\documentstyle [a4,12pt]{article}
\newcommand{\etal}{{et al.}~}
\newcommand{\done}{\delta^{(1)}}
\newcommand{\p}{\partial}
\newcommand{\f}{\frac}
\newcommand{\ap}{\approx}
\newcommand{\Om}{\Omega}
\newcommand{\s}{\sigma}
\newcommand{\al}{\alpha}
\newcommand{\fd}{\tilde{\delta}}

\newcommand{\bfx}{{\bf x}}
\newcommand{\bfy}{{\bf y}}
\newcommand{\bfk}{{\bf k}}
\newcommand{\bfv}{{\bf v}}
\newcommand{\bfq}{{\bf q}}

\newcommand{\bc}{\begin{center}}
\newcommand{\be}{\begin{equation}}
\newcommand{\ee}{\end{equation}}
\newcommand{\ec}{\end{center}}
\newcommand{\lan}{\langle}
\newcommand{\ran}{\rangle}

\title{{\huge {\bf KURTOSIS AND LARGE--SCALE STRUCTURE}}\\}
\author{ {\bf Paolo CATELAN}$^{\,1}$ and
{\bf Lauro MOSCARDINI}$^{\,2}$ \\ \\
$^{1\,}$ {\it SISSA--International School for Advanced Studies} \\
{\it via Beirut 2--4, I--34013 Trieste, Italy} \\ \\
$^{2\,}${\it Dipartimento di Astronomia, Universit\`a di Padova} \\
{\it vicolo dell'Osservatorio 5, I--35122 Padova, Italy} \\ }
\date{}
\begin{document}
\maketitle
\vspace{2cm}
\bc
{\it Astrophysical Journal Letters}, submitted
\ec
\vspace{2cm}
\bc
{\it SISSA Ref. 127/93/A}
\ec
\newpage
\vspace{1cm}
\bc
\section*{Abstract}
\ec
We discuss the non--linear growth of the excess kurtosis parameter of the
smoothed density fluctuation field $\delta$,
$S_4\equiv[\lan\delta^{\,4}\ran-3\lan\delta^{\,2}\ran^2]/
\lan\delta^{\,2}\ran^3$ in an Einstein--de Sitter universe. We assume Gaussian
primordial density fluctuations with scale--free power spectrum $P(k)\propto
k^{\,n}$ and analyze the dependence of $S_4$ on primordial spectral index $n$,
after smoothing with a Gaussian filter. As already known for the skewness ratio
$S_3$, the kurtosis parameter is a {\it decreasing function} of $n$, both in
exact perturbative theory and in the Zel'dovich approximation. The parameter
$S_4$ provides a powerful statistics to test different cosmological scenarios.

\vspace{0.5cm}
\noindent{\em Subject headings:} Galaxies: clustering -- large--scale
structure of the Universe
\vspace{1.0cm}

\section{Introduction}

The study of the statistical distribution of the matter in the universe may be
a way to address fundamental issues such as the origin and the formation of
structures on large scales.

The simplest and most usually accepted hypothesis, supported by the
inflationary model, is that the very early distribution is Gaussian. In such a
case, the connected $N$--point correlations (and the $N$--th order connected
moments) with $N>2$ are zero. However, even if the primordial fluctuations
$\delta$ are Gaussian, the non--linear time evolution will ensure that the mass
density fluctuations become highly non--Gaussian (Peebles 1980; Fry 1984). It
is important to understand the nature of the higher moments of the mass density
induced by gravity in order to distinguish their effects from those of possible
$primordial\,$ non--Gaussian fluctuations: late--time phase transitions, cosmic
string models and global textures are indeed models whose statistics may not be
described by a Gaussian distribution (see e.g. Vilenkin 1985; Turok 1989;
Scherrer \& Bertschinger 1991). Moreover, variations of the inflationary model
which lead to non--Gaussian primordial fluctuations have been recently
discussed (see e.g. Salopek 1992).

A powerful method to distinguish if the non--Gaussian nature of the matter
distribution is intrinsic, is to analyze the growth of higher moments, like the
skewness or the kurtosis (third and fourth connected moments, respectively) of
the density fluctuation field. Peebles (1980) first showed that gravity
induces, for an unsmoothed initial Gaussian density field, a skewness ratio
$S_3\equiv\lan\delta^{\,3}\ran/\lan\delta^{\,2}\ran^2=34/7$, for any primordial
power spectrum. Juszkiewicz, Bouchet, \& Colombi (1993) find that the filtering
operation actually introduces a dependence of $S_3$ on the primordial spectral
index $n$; in particular for scale--free spectra, $S_3=34/7-(n+3)$ for a
top--hat window function, and a decreasing trend with $n$ is recovered also for
a Gaussian filter. Coles \etal (1993) investigate, using $N$--body simulations,
the growth of the skewness ratio to test the hypothesis of Gaussian primordial
density fluctuations against possible alternatives. Analytical expressions for
$S_3$ arising from non--linear evolution in perturbation theory have been
worked out for arbitrary non--Gaussian models by Fry \& Scherrer (1993) and
Catelan \& Moscardini (1993). Finally, higher order moments are known to depend
extremely weakly on the density parameter $\Om$ (see Martel \& Freudling 1991;
Bouchet \etal 1992; Bernardeu 1992).

Here we analyse the dependence of the induced--by--gravity excess kurtosis of
the density field, smoothed with a Gaussian filter, namely the parameter
$S_4\equiv[\lan\delta^{\,4}\ran-3\lan\delta^{\,2}\ran^2]/
\lan\delta^{\,2}\ran^3$, on an initial (scale--free) power spectrum $P(k)$. To
do this, we take advantage of the exact perturbative technique (Fry 1984;
Goroff \etal 1986) and the Zel'dovich approximation (see Grinstein \& Wise
1987). The kurtosis describes features such as sharpness or stretchiness of the
mass distribution and the extent of its rare--event tail. Moreover, it is
possibly related to the $initial\,$ sign of the skewness -- as predicted in
some non--Gaussian models (see e.g. Luo \& Schramm 1993) -- which is important
for the final galaxy clustering pattern (Moscardini \etal 1991; Messina \etal
1992; Weinberg \& Cole 1992).

The layout of this {\it Letter} is as follows: in Section 2 we review the exact
perturbative theory and the Zel'dovich approximation; in Section 3 we discuss
the induced--by--gravity kurtosis parameter $S_4$ of an initial Gaussian
density field and its dependence on the primordial spectral index; we state the
main conclusions in Section 4.
\section{Non--Linear Time Evolution}
We assume that present--day structures in the universe formed by gravitational
instability from Gaussian fluctuations $\delta$ in a pressureless fluid with
matter density $\rho=\rho_b[1+\delta]$, where $\rho_b$ is the background mean
density. The density fluctuation field $\delta$ may be written as a Fourier
integral,
$
\delta(\bfx,t)=(2\pi)^{-3}\int d\bfk\,\fd(\bfk,t)\,
{\rm e}^{\,i\,\bfk\cdot\bfx}\,,
$
where $\bfx$ and $\bfk$ are the {\it comoving} Eulerian coordinate and
wavevector respectively, $t$ is the cosmic time. The power spectrum $P(k)$,
defined as the 2--point correlation function in Fourier space (at a given
time),
$
\lan\fd(\bfk_1)\,\fd(\bfk_2)\ran\equiv(2\pi)^3\,
\delta_D(\bfk_1+\bfk_2)\,P(k_1)\,,
$
fully determines the statistics of the primordial Gaussian density field, whose
variance is given by the expression
$\s^{\,2}=(1/2\pi^2)\int_0^{\infty}dk\,k^2\,P(k)$.

To bridge the gap between observational data and theory, it is {\it necessary}
to filter the fluctuation field $\delta$ by means of a {\it window function}
$W_R\,$,
$
\delta_R(\bfx,t) \equiv
\int d\bfy\,\delta(\bfy,t)\,W_R(|\bfx - \bfy|)\;.
$
In the following we will adopt a Gaussian window function. The mass variance on
scale $R$, $\s_R^2\,$, is related to the primordial spectrum by
$\s_R^{\,2}=(1/2\pi^2)\int_0^{\infty}dk\,k^2\,P(k)\,[\widetilde{W}_R(k)]^2\,$,
where $\widetilde{W}_R(k)$ is the Fourier transform of $W_R(x)\,$.
\subsection{Equations of Motion: Perturbative Theory}
The time evolution equations for the matter density fluctuation
$\delta(\bfx,t)$ and the peculiar velocity field $\bfv(\bfx, t)$ are the
Poisson equation, the Euler equation and the continuity equation, i.e.
\be
\nabla^2\Phi=4\pi\,G\,\rho_b\,a^2\,\delta\;,
\ee
\be
\p_{\circ}\bfv+\f{1}{a}\,(\bfv\cdot\nabla)\,\bfv+
\f{\dot{a}}{a}\,\bfv=-\f{1}{a}\,\nabla\Phi\;,
\ee
\be
\p_{\circ}\delta+\f{1}{a}\,\nabla\cdot(1+\delta)\,\bfv=0\;.
\ee
Here $\p_{\circ}\equiv\p/\p t$ and spatial derivatives are with respect to
$\bfx$. We analyze these equations assuming an Einstein--de Sitter universe
with vanishing cosmological constant. In such a model, the scale factor $a$ is
proportional to $t^{\,2/3}$ during the matter dominated epoch, and the
adiabatic expansion implies that $6\pi\,G\rho_b\,t^2 =1\,$. The quantity $\Phi$
is the Newtonian gravitational potential. Combining the divergence of the Euler
equation with the continuity equation, a second order differential equation for
the density contrast $\delta$ may be introduced (Peebles 1980)
\be
\p_{\circ}^2\delta+2\,\f{\dot{a}}{a}\,\p_{\circ}\delta-4\pi\,G\rho_b\,\delta=
4\pi\,G\,\rho_b\,\delta^2+\f{1}{a^2}
\,\left[\p_{\al}\delta\,\p_{\al}\Phi+\p_{\al}\p_{\beta}\,
(1+\delta)v^{\al}v^{\beta}\right]\;.
\ee
The first term in the r.h.s. of Eq.(4) corresponds to the spherical collapse of
an isolated proto--object, while the ``geometrical'' term in square brackets
describes tidal and shear effects. The linear approximation, adequate when
$\delta^{\,2} \ll 1$, corresponds to dropping the r.h.s. of Eq.(4). The first
order solution has the well--known self--similar form (considering only the
growing mode)
\be
\done(\bfx,t)=D(t)\,\delta_1(\bfx)\;,
\ee
where $D(t)\propto a(t)$ is the time growth factor of the mass fluctuations.
Higher order approximations of the solution of Eq.(4) may be recovered if one
expands the mass density fluctuation field $\delta(\bfx, t)$ about the
background solution $\delta=0$ (and $\bfv={\bf 0})$, namely
$\delta=\sum_n\delta^{(n)}$ with $\delta^{(n)}=O(\delta_1^n)$, then solving the
differential equation for any $\delta^{(n)}$ (Peebles 1980; Fry 1984). The
perturbative expansion for $\delta$ is (e.g. Goroff \etal 1986)
\be
\delta(\bfx,t)=\sum_{n=1}^{\infty}[D(t)]^n\,\delta_n(\bfx)\;.
\ee
The first term of the expansion corresponds to the linear approximation. Second
and third order solutions have been discussed by Peebles (1980) and Fry (1984)
respectively. We see that the scale factor $D(t)$ acts as a coupling constant
in this perturbative approach, since $\delta^{(n)} \propto D^n$.

Here we review the exact perturbative technique to solve approximately the
equations of motion of a pressureless gravitational fluid up to third order in
the density fluctuation field. In particular, we will use the third order
solution to compute the fourth order moment (namely the kurtosis parameter
$S_4$). We use explicitly the same notation of Fry (1984).
\bc
{\it (i) Second Order Density Solution}
\ec
The second order contribution $\delta^{(2)}$ is related to the second order
gravitational potential $\Phi^{(2)}$ by the Poisson equation
$\nabla^2\Phi^{(2)}=4\pi G a^2\rho_b\,\delta^{(2)}\,$; it is solution of the
differential equation
$$
\p_{\circ}^2\delta^{(2)}+2\dot{a} a^{-1}\p_{\circ}\delta^{(2)} -
4\pi G \rho_b\,\delta^{(2)}=
\left[ 4\pi G\rho_b + \left(\f{\dot{D}}{D}\right)^2\,\right]\delta^{(1)\,2}\,+
$$
\be
+\,\left[ 4\pi G\rho_b + 2\left( \f{\dot{D}}{D} \right)^2\,\right]
\p_{\alpha}\delta^{(1)}\, \p_{\alpha}\triangle^{(1)} +
\left( \f{\dot{D}}{D} \right)^2
\p_{\alpha}\p_{\beta}\triangle^{(1)}\,
\p_{\alpha}\p_{\beta}\triangle^{(1)}\;,
\ee
where $\triangle^{(1)}$ is the rescaled linear gravitational potential defined
by $\triangle^{(1)}\equiv\Phi^{(1)}/4\pi G\rho_b\,a^2$, for which
$\nabla^2\triangle^{(1)}=\delta^{(1)}$. Each side of Eq.(7) is homogeneous in
powers of $t\,$. For an Einstein--de Sitter universe, noting that $D^2
\propto\delta^{(2)}\equiv D^2(t)\,\delta_2(\bfx)$, the solution of Eq.(7) is
given by (Peebles 1980, \S18)
\be
\delta^{(2)}\, =\,
\f{5}{7}\,\delta^{(1)2} +
\p_{\alpha}\delta^{(1)}\, \p_{\alpha}\triangle^{(1)} +
\f{2}{7}\,\p_{\alpha}\p_{\beta}\triangle^{(1)}\,
\p_{\alpha}\p_{\beta}\triangle^{(1)}\;.
\ee
Note that $\lan \delta^{(2)}\ran=0$, i.e. mass is conserved to second order. We
see from Eq.(8) that the behaviour of $\delta$ at second order is non--local:
the mass fluctuation at the position $\bfx$ depends on initial perturbations at
other positions via $\triangle^{(1)}\,$. Physically this means that, unlike the
linear local case, when density fluctuations grow in amplitude their spatial
dependence, in comoving coordinates, changes. Thus, the gravitational field
changes direction and particles are not accelerated in a fixed direction, as it
occurs in linear regime. The last term in right--hand side of Eq.(8)
corresponds to the velocity shear contribution.

We can obtain the Fourier transform $\fd^{(2)}$ directly from the differential
equation (8). Explicitly (time dependence is understood),
\be
\fd^{(2)}(\bfk) = \int \f{d\bfk'}{(2\pi)^3}\,
\,J^{(2)}(\bfk',\bfk-\bfk')\,\,\fd^{(1)}(\bfk')
\,\fd^{(1)}(\bfk - \bfk')\;,
\ee
where we have defined the kernel
\be
J^{(2)}(\bfk,\bfk') \equiv
\f{5}{7} + \f{\bfk\cdot\bfk'}{k'^{\,2}} +
\f{2}{7}\left(\f{\bfk\cdot\bfk'}{k\,k'} \right)^2\;.
\ee
\bc
{\it (ii) Third Order Density Solution}
\ec
The third order approximation $\delta^{(3)}$ is solution of the differential
equation
$$
\p_{\circ}^2\delta^{(3)}+2\dot{a} a^{-1}\p_{\circ}\delta^{(3)} -
4\pi G \rho_b\,\delta^{(3)}
= 8\pi G \rho_b\,\delta^{(1)}\,\delta^{(2)}+
$$
\be
+a^{-2}\left(\p_{\alpha}\delta^{(1)}\,\p_{\alpha}\Phi^{(2)}+
\p_{\alpha}\delta^{(2)}\,\p_{\alpha}\Phi^{(1)}\right) +
a^{-2}\,\p_{\alpha}\p_{\beta}\left(2\,v^{(1)\,\alpha}v^{(2)\beta} +
\delta^{(1)}\,v^{(1)\alpha}\,v^{(1)\beta}\right)\;.
\ee
In this case, working directly in Fourier space is much simpler. Since
$\delta^{(3)}\propto D^3\propto t^2$, we have $\p_{\circ}^2\fd^{(3)}+2\dot{a}
a^{-1}\p_{\circ}\fd^{(3)} - 4\pi G \rho_b\,\fd^{(3)}=4\,\fd^{(3)}/t^2\,$; by
using the second order results it is not difficult to show that (Fry 1984)
\be
\fd^{(3)}(\bfk)=
\int \f{d\bfk_1\,d\bfk_2\,d\bfk_3}{(2\pi)^6}
\,\,\delta_D(\bfk_1+\bfk_2+\bfk_3 - \bfk)
\,\,J^{(3)}(\bfk_1, \bfk_2, \bfk_3)
\,\,\fd^{(1)}(\bfk_1)\,\fd^{(1)}(\bfk_2)
\,\fd^{(1)}(\bfk_3)\;,
\ee
where the third order kernel is
$$
J^{(3)}(\bfk_1,\bfk_2,\bfk_3)\equiv J^{(2)}(\bfk_2, \bfk_3)
\left[\f{1}{3}+
\f{1}{3}\,\f{\bfk_1\cdot(\bfk_2+\bfk_3)}{(\bfk_2+\bfk_3)^2} +
\f{4}{9}\, \f{\bfk\cdot\bfk_1}{k_1^{\,2}} \,
\f{\bfk\cdot(\bfk_2+\bfk_3)}{(\bfk_2+\bfk_3)^2}
\right]+
$$
\be
- \f{2}{9}\, \f{\bfk\cdot\bfk_1}{k_1^{\,2}}
\,\f{\bfk\cdot(\bfk_2+\bfk_3)}{(\bfk_2+\bfk_3)^2}
\,\f{(\bfk_2+\bfk_3)\cdot\bfk_3}{k_3^{\,2}}+
\f{1}{9}\, \f{\bfk\cdot\bfk_2}{k_2^{\,2}} \,\f{\bfk\cdot\bfk_3}{k_3^{\,2}}
\;.
\ee
It is clear from Eq.(11) that, in order to derive the solution $\delta^{(3)}$,
we need to know explicitly the second order velocity $\bfv^{(2)}$. This is
given by
\be
\bfv^{(2)}=a\,\f{\dot{D}}{D}\,
\left[\delta^{(1)}\,\nabla\triangle^{(1)}-2\,\nabla\triangle^{(2)}\right]\;,
\ee
where $\triangle^{(2)}\equiv\Phi^{(2)}/4\pi G\rho_b\,a^2$ is the rescaled
second order gravitational potential. [In the solution (14) we neglect an
additive homogeneous term whose divergence is zero.] In an Einstein--de Sitter
universe $\bfv^{(2)}\sim t$, slower than $\delta^{(2)}\sim t^{\,4/3}$. We
stress the fact that $\bfv^{(2)}$ is {\it not} parallel to the second order
acceleration [$\propto -\nabla\triangle^{(2)}\,$]: this is a consequence of
non--locality. Finally, we note that $\bfv^{(2)}$ is known only once
$\delta^{(2)}$ is known.
\bc
{\it (iii) General Solution}
\ec
As Goroff \etal (1986) have shown, the (Fourier transformed) $n$--th order mass
fluctuation term may be represented in integral form as
\be
\fd_n(\bfk)=\left\{\prod_{h=1}^n\int\f{d\bfk_h}{(2\pi)^3}\,\fd_1(\bfk_h)\,
\right\}
\,\Bigl[(2\pi)^3\delta_D(\sum_{j=1}^n\bfk_j-\bfk)\Bigr]\,J^{(n)}
(\bfk_1,\dots,\bfk_n)\;.
\ee
The presence of the Dirac delta function comes from momentum conservation in
Fourier space. The kernels $J^{(n)}$ are symmetric homogeneous (with degree 0)
functions of the wavevectors $\bfk_1,\ldots,\bfk_n\,$, describing the effects
of non--linear collapse (tidal and shear forces). In general $J^{(n)}$ are very
complicated for $n>3$. [A discussion of the properties of the kernels $J^{(n)}$
is given in Wise (1988). Explicit recursion relations with their Feynman
diagrammatic representation are given by Goroff \etal (1986) and Wise (1988).]
\subsection{Zel'dovich Approximation}
In the {\it Zel'dovich approximation} (Zel'dovich 1970) the motion of particles
from the initial comoving (Lagrangian) positions $\bfq\,$ is approximated by
straight paths. The Eulerian position at time $t$ is then given by the
uniformly accelerated motion
\be
\bfx[\bfq,t]=\bfq+D(t)\,{\bf S}(\bfq)\;,
\ee
where $D(t)$ is the growth factor of linear density perturbations and ${\bf
S}(\bfq)$ is the displacement vector related to the primordial velocity field.
The Zel'dovich approximation provides an exact solution of the equations of
motion for one--dimensional perturbations, and reduces to the linear
approximation when $\delta$ and $\bfv$ are small. In general, the Zel'dovich
approximation is not an exact solution of the equations of motion, in that in a
finite time particles converge into singular regions of infinite density
(caustics), and the map in Eq.(16) becomes multi--valued. The treatment of
these formal singularities is the main problem to solve during the highly
non--linear stage of structure formation. Smoothing on a suitable scale $R$
partially solves this problem. Grinstein \& Wise (1987) give an Eulerian
representation of the Zel'dovich approximation by a diagrammatic perturbative
approach similar to that of the previous section. They showed that the $n$--th
order perturbative corrections $\delta_n(\bfx)$, when the density fluctuation
field $\delta$ is evolved according to the Zel'dovich approximation, are such
that
\be
\delta(\bfx,t)=\sum_{n=1}^{\infty}\f{(-1)^n}{n!}\,[D(t)]^n
\sum_{[h_n]=1}^3\f{\p}{\p x_{h_1}}\cdots\f{\p}{\p x_{h_n}}\,
\Bigl[S_{h_1}\cdots S_{h_n}\Bigr]\;.
\ee
Here $\sum_{[h_n]}\equiv\sum_{h_1}\cdots\sum_{h_n}$. Note that the first term
recovers the linear approximation, in that ${\bf S}=\bfv_1$, where
$\delta_1(\bfx)=-\nabla\cdot\bfv_1\,$. The perturbative expansion for $\delta$
of Eq.(17) simply corresponds to calculate different symmetric kernels
$J_{ZA}^{(n)}$. These can be written in the following compact form
\be
J_{ZA}^{(n)}(\bfk_1,\ldots,\bfk_n)=
\f{1}{n!}\,\prod_{h=1}^n\,\f{\bfk\cdot\bfk_h}{k_h^{\,2}}\;,
\ee
where $\bfk\equiv\sum_{h=1}^n\bfk_h\,$. Note that, unlike the perturbative
case, the kernels $J_{ZA}^{(n)}$ are manifestly symmetric by construction.
\section{\bf Kurtosis of the Density Field}
In this section, we compute the induced--by--gravity kurtosis parameter $S_4$
of an initial Gaussian density field in a flat universe. The lowest order
non--zero contribution to $S_4$ is
\be
\lan\delta^{(1)\,2}\ran^3\,\,S_4\, =\, 6\,\lan
\,\delta^{(1)\,2}\,\delta^{(2)\,2}\,\ran
\,+\,4\,\lan\,\delta^{(1)\,3}\,\delta^{(3)}\,\ran\;.
\ee
{}From Eqs.(10) and (13), after tedious but straightforward algebra, one
finally
gets the integral expression of the kurtosis ratio,
$$
S_4 = \f{24}{\sigma^{\,6}}
\int \f{d\bfk_1\,d\bfk_2\,d\bfk_3}{(2\pi)^9}\,\,P(k_1)\,P(k_2)\,
\left[P(k_3)\,J^{(3)}(\bfk_1, \bfk_2, \bfk_3)\,\right.+
$$
\be
\left.+2\,J^{(2)}(-\bfk_2, \bfk_2+\bfk_3)\,
J^{(2)}(\bfk_1, \bfk_2 + \bfk_3)\,P(|\bfk_2+\bfk_3|)\right] \;.
\ee
The expression in the Zel'dovich approximation is obtained by substituting the
kernels $J^{(n)}$ with the corresponding $J^{(n)}_{ZA}$ in the previous
relation. The angular integrations in Eq.(20) may be analytically performed.
One obtains $S_4=60,712/1,323\ap 45.9$ in the perturbative case (Fry 1984;
Bernardeu 1992) and $S_4=88/3\ap 29.3$ in the Zel'dovich approximation. These
unsmoothed--case results are independent of the primordial spectral index, as
is the skewness ratio $S_3$.

The kurtosis of the smoothed density field $\delta_R$ in the exact perturbative
case is
$$
S_4(R) = \f{24}{\sigma_R^{\,6}}
\int \f{d\bfk_1\,d\bfk_2\,d\bfk_3}{(2\pi)^9}\,\,
\widetilde{W}_R(k_1)\,\widetilde{W}_R(k_2)\,\widetilde{W}_R(k_3)\,
\widetilde{W}_R(|\bfk_1+\bfk_2+\bfk_3|)\,\times
$$
\be
P(k_1)\,P(k_2)\,
\left[P(k_3)\,J^{(3)}(\bfk_1, \bfk_2, \bfk_3)\,+\,
2\,J^{(2)}(-\bfk_2, \bfk_2+\bfk_3)\,
J^{(2)}(\bfk_1, \bfk_2 + \bfk_3)\,P(|\bfk_2+\bfk_3|)\right]\;;
\ee
again the substitution $J^{(n)} \rightarrow J^{(n)}_{ZA}$ leads to the
corresponding expression in the Zel'dovich approximation. Note that $P(k)$
completely describes the process of growth of mass fluctuations from Gaussian
initial perturbations.
\section{Discussion and Conclusions}
We calculate the previous integrals, by an Adaptive Multidimensional Monte
Carlo Integration subroutine, for scale--free power spectra $P(k) \propto
k^{\,n}$ with $n$ in the range $-3\le n \le 1$. Due to the assumed primordial
scale--invariance, $S_4$ only depends on the primordial spectral index $n$, and
not on the scale $R$. In Figure 1, we plot the kurtosis ratio $S_4$ of the
Gaussian--smoothed density field versus the spectral index $n$, for both the
perturbative and Zel'dovich approximations.

It clearly appears that $S_4\,$ $strongly\,$ depends on the primordial spectral
index $n$. In particular the kurtosis parameter is a $decreasing$ function of
$n\,$. This is also confirmed by Bernardeau (1993), who, applying the exact
perturbative technique, finds a similar trend in the case of top--hat
filtering. Anticorrelation with the amount of small--scale power has also been
found for the skewness parameter $S_3$ (Fry 1984; Juszkiewicz, Bouchet, \&
Colombi 1993), and it seems therefore a general property of higher order
moments of the smoothed matter distribution: larger values of $n$ correspond to
higher power on small scales, where the filtering operation acts.

Furthermore, the Zel'dovich approximation underestimates the
induced--by--gravity $S_4$ w.r.t. the rigorous perturbative one, as already
noted by Grinstein \& Wise 1987 (although in the framework of the ``standard''
cold dark matter model). This is hardly surprising, since the Zel'dovich
approximation fails to fully describe the gravitational effects causing
particle trajectories to depart from their original directions. It results that
the Zel'dovich approximation makes higher orders in the perturbative series for
$\delta$ smaller on {\it large} scales than they actually are (see e.g. Wise
1988).

We stress that, due to the importance of the smoothing procedure, but also to
its arbitrariness, one must be cautious about making quantitative comparisons
between our results and both observational and $N$--body data. A Gaussian
filter is used for instance  by Saunders \etal (1991), who however estimated
only the skewness in the QDOT--$IRAS$ catalog. The only observational estimates
up--to--now of the kurtosis of the galaxy distribution are obtained by
counts--in--cells, corresponding to a top--hat filtering (Gazta\~naga 1992;
Bouchet \etal 1993; Fry \& Gazta\~naga 1993); a similar method is applied by
Lahav \etal (1993) and Lucchin \etal (1993) in $N$--body simulations with
scale--free power spectra. Our results can be used to constrain both the
probability distribution and the power spectrum of primordial density
fluctuations and thus help to select the most reliable mechanism for the
generation of large--scale structures.

\section*{Acknowledgments}
Sabino Matarrese is warmly thanked for addressing our attention to the
cosmological density kurtosis, and for carefully reading the manuscript. PC
acknowledges Robert Scherrer for his kind hospitality at the Ohio State
University, Columbus, where part of this work begun. This work has been
supported by Italian MURST.

\newpage

\parindent 0pt
\section*{References}

\begin{trivlist}

\item[] Bernardeau, F. 1992, ApJ, 392, 1

\item[] Bernardeau, F. 1993, in preparation

\item[] Bouchet, F.R., Juszkiewicz, R., Colombi, S., \& Pellat, R. 1992, ApJ,
	394, L5

\item[] Bouchet, F.R., Strauss, M., Davis, M., Fisher, K.B., Yahil, A., \&
        Huchra, J.P. 1993, ApJ, in press

\item[] Catelan, P., \& Moscardini, L. 1993, in preparation

\item[] Coles, P., Moscardini, L., Lucchin, F., Matarrese, S., \& Messina,
        A. 1993, MNRAS, in press

\item[] Fry, J.N. 1984, ApJ, 279, 499

\item[] Fry, J.N., \& Gazta\~naga, E. 1993, preprint

\item[] Fry, J.N., \& Scherrer, R.J. 1993, preprint

\item[] Gazta\~naga, E. 1992,  ApJ, 398, L17

\item[] Goroff, M.H., Grinstein, B., Rey, S.--J., \& Wise, M.B. 1986, ApJ,
	311, 6

\item[] Grinstein, B., \& Wise, M.B. 1987, ApJ, 314, 448

\item[] Juszkiewicz, R., Bouchet, F.R., \& Colombi, S. 1993, ApJ, 412, L9

\item[] Lahav, O., Itoh, M., Inagaki, S., \& Suto, Y. 1993,  ApJ, 402, 387

\item[] Lucchin, F., Matarrese, S., Melott, A.L., \& Moscardini, L. 1993,
        ApJ, submitted

\item[] Luo, X.C., \& Schramm, D.N. 1993, ApJ, 405, 30

\item[] Martel, H., \& Freudling, W. 1991, ApJ, 371, 1

\item[] Messina, A., Lucchin, F., Matarrese, S., \& Moscardini, L. 1992,
        Astroparticle Phys., 1, 99

\item[] Moscardini, L., Matarrese, S., Lucchin, F., \& Messina, A. 1991, MNRAS,
	248, 424

\item[] Peebles, P.J.E. 1980, The Large Scale Structure of the Universe
        (Princeton: Princeton Univ. Press)

\item[] Salopek, D.S. 1992, Phys. Rev., D45, 1135

\item[] Saunders, W., \etal 1991, Nature, 349, 32

\item[] Scherrer, R.J., \& Bertschinger, E. 1991, ApJ, 381, 349

\item[] Turok, N. 1989, Phys. Rev. Lett., 63, 2625

\item[] Vilenkin, A. 1985, Phys. Rep., 121, 263

\item[] Weinberg, D.H., \& Cole, S. 1992, MNRAS, 259, 652

\item[] Wise, M.B. 1988, in The Early Universe, ed. W.G. Unruh \&
        G.W. Semenoff, Reidel Pub., p.215

\item[] Zel'dovich, Ya.B. 1970, A\&A, 5, 160
\end{trivlist}

\newpage

\section*{\center Figure caption}

{\bf Figure 1.} The kurtosis ratio $S_4$ of the density field for power--law
spectra $P(k)\propto k^{\,n}$ and Gaussian filter versus the primordial
spectral index $n$, for both the perturbative (squares and solid line) and
Zel'dovich approximations (triangles and dotted line). Note that the values at
$n=-3$ correspond to the unsmoothed cases. Error bars refer to the associated
uncertainty estimate from the Monte Carlo Integration.

\end{document}